\documentclass[twocolumn, titlepage, superscriptaddress, amsmath, amssymb, 10pt, prb]{revtex4-2}

\usepackage{mathtools}        
\usepackage{bbm}
\usepackage{float}            
\usepackage{graphicx}         
\usepackage{color}
\usepackage[dvipsnames]{xcolor}
\usepackage[export]{adjustbox}
\usepackage[caption=false]{subfig}
\usepackage{physics}
\usepackage{epstopdf}
\usepackage{comment}
\usepackage{hyperref}
\usepackage{array}
\hypersetup{
	linkcolor=blue,
	citecolor=blue,
	filecolor=blue,
	urlcolor=blue,
	colorlinks=true
}

\usepackage{hhline}
\usepackage{multirow}

\usepackage{soul} 

\renewcommand{\vb}[1]{\ensuremath{\boldsymbol{\mathbf{#1}}}}

\newcommand{\UIUCPHYS}[0]{Department of Physics and Institute of Condensed Matter Theory, University of Illinois at Urbana-Champaign, Urbana, IL 61801, USA}

\begin{document}

\title{Crystalline-electromagnetic responses of higher order topological semimetals}

\author{Mark R. Hirsbrunner}\email[Corresponding author: ]{hrsbrnn2@illinois.edu}\affiliation{\UIUCPHYS}
\author{Alexander D. Gray}\affiliation{\UIUCPHYS}
\author{Taylor L. Hughes}\affiliation{\UIUCPHYS}

\begin{abstract}
Previous work has shown that time-reversal symmetric Weyl semimetals with a quadrupolar arrangement of first-order Weyl nodes exhibit a mixed crystalline-electromagnetic response. For systems with higher order Weyl nodes, which are attached to both surface and hinge Fermi arcs, additional phenomena appear on surfaces of codimension $n>1$, such as electromagnetic responses of the hinges. Here we construct a model possessing a quadrupole of higher order Weyl nodes to study the interplay between higher order topology and mixed crystalline-electromagnetic responses. We show that the higher order nature of the Weyl nodes yields a dipole of Dirac nodes on certain surfaces, leading to a mixed crystalline-electromagnetic \textit{surface} response that binds charge to dislocations and momentum-density to magnetic fields. In addition, we show that the model possesses a bulk quadrupole moment of crystal-momentum that provides a link between the bulk and surface responses of the system.
\end{abstract}

\maketitle

\section{Introduction}
Topological semimetals (TSMs) possess quasi-topological terms in their bulk electromagnetic responses that are governed by the configuration of their nodal points or lines in momentum space~\cite{yan_topological_2017, armitage_weyl_2018, bernevig_recent_2018, lv_experimental_2021, gao_topological_2019, hu_transport_2019, wang_quantum_2017, burkov_weyl_metals_2018}. In particular, the responses of point node TSMs are proportional to the chirality-weighted momentum space multipole moments of the nodal points, i.e.,  monomials of their momentum-space location weighted by their chirality or helicity. For example, in the simplest case of a time-reversal breaking Weyl semimetal (WSM) with two nodes, the magnitude of the bulk anomalous Hall conductivity is proportional to the dipole moment of the Weyl nodes in momentum space~\cite{burkov_weyl_2011,wan2011WSM, zyuzin_topological_2012, ramamurthy_patterns_2015}. Additionally, these bulk responses are often necessary to compensate for anomalous surface states, such as chiral Fermi arcs in time-reversal breaking WSMs~\cite{wan2011WSM}.

In recent years the field of TSMs has grown to include higher order TSMs (HOTSMs) that are characterized by spectral features and other phenomena on surfaces of codimension $n>1$. The nodal points of HOTSMs differ from conventional TSMs in that they are attached to both surface and hinge Fermi arcs. Heuristically such a node separates gapped momentum space planes that differ in both Chern number \emph{and} some form of 2D higher order topology. The family of HOTSMs is quite diverse, including higher order analogs of Dirac and Weyl semimetals~\cite{lin_topological_2018, fangClassificationDiracPoints2021a, wieder2020strong,PhysRevLett.120.026801, zengHigherorderTopologicalInsulators2020, wangHigherOrderTopologyMonopole2019, szaboDirtyHigherorderDirac2020, liReducingElectronicTransport2020,ezawaSecondorderTopologicalInsulators2019, calugaruHigherorderTopologicalPhases2019a, ghorashiHigherOrderWeylSemimetals2020, wangHigherOrderWeylSemimetals2020, tanakaRotoinversionsymmetricBulkhingeCorrespondence2022, ruiIntertwinedWeylPhases2022a, songSquarerootHigherorderWeyl2022}, nodal line semimetals~\cite{wangBoundaryCriticalityMathcalPT2020}, nodal superconductors~\cite{ghorashi2019second, PhysRevB.103.184510, wuNodalHigherorderTopological2022a, simonHigherorderTopologicalSemimetals2022}, non-Hermitian TSMs~\cite{ghorashi2021nonDirac,ghorashi2021nonWeyl,liu2021higher, bidNonHermitianHigherOrderWeyl2022}, and periodically driven Floquet TSMs~\cite{zhuFloquetHigherorderWeyl2021, wuHybridorderTopologicalOddparity2022, wangFloquetWeylSemimetal2022, ghoshHingemodeDynamicsPeriodically2022, duWeylNodesHigherorder2022}. In some instances, HOTSMs possess additional boundary states and/or electromagnetic responses beyond first-order TSMs. For example, second order WSMs exhibit both surface Fermi arcs and hinge states that generate competing surface and hinge responses~\cite{ghorashiHigherOrderWeylSemimetals2020} in which the bulk charge bound to a magnetic flux (via the anomalous Hall effect) is constrained by the charge bound to hinges parallel to the flux.  Similarly, conventional type-I Dirac semimetals (DSMs) have a bulk spin-Hall-like response determined by the momentum-space dipole moment of the Dirac nodes~\cite{ramamurthy_patterns_2015}, while some higher order DSMs also possess a bulk electric quadrupole moment that generates a surface polarization response~\cite{lin_topological_2018}.

In parallel to these developments of HOTSMs, recent studies have shown that TSMs can exhibit mixed crystalline-electromagnetic responses in addition to purely electromagnetic responses. These mixed crystalline-electromagnetic responses are often probed by subjecting systems to dislocation and disclination defects~\cite{teoTopologicalDefectsSymmetryProtected2017a}. TSMs typically possess interesting response phenomena to such defects because the TSM nodal surfaces are protected by translation symmetry and, in some cases, rotation symmetries~\cite{you_response_2016, torsional_chiral_2016, dislocation_defect_2020, laurila_torsional_2020, dubinkinHigherRankChiral2021, gioia_unquantized_2021, wangEmergentAnomaliesGeneralized2021, nissinenTopologicalPolarizationDual2021, gao_chiral_2021, you_fracton_2022, dubinkin_in_prep, amitani_torsion_2023}. For example, time-reversal symmetric WSMs with a quadrupole arrangement of Weyl nodes in momentum space have electric charge bound to screw dislocations and crystal momentum bound to magnetic flux~\cite{dubinkinHigherRankChiral2021, dubinkin_in_prep, gioia_unquantized_2021}.

Motivated by these unusual electromagnetic responses, here we take the first steps toward understanding the mixed crystalline-electromagnetic responses of higher order TSMs. In Section~\ref{sec:model} we introduce a model of a TSM with a quadrupole arrangement of higher order Weyl nodes and characterize its topological features. In Section~\ref{sec:mixed_responses} we show that this model possesses a rank-2 mixed crystalline-electromagnetic response similar to that found in Refs.~\onlinecite{dubinkinHigherRankChiral2021, gioia_unquantized_2021} for quadrupolar arrangements of first-order Weyl nodes. Furthermore, we demonstrate that the higher order nature of the Weyl nodes in our model leads to an additional \emph{surface} crystalline-electromagnetic response arising from the presence of a dipole of surface Dirac nodes. In Section~\ref{sec:momentum_quadrupole_moment} we show that this model can possess a bulk quadrupole moment of equilibrium crystal momentum. We show that the magnitude of this quadrupole moment of momentum is determined by both the momentum-space quadrupole moment of the bulk Weyl nodes and the momentum-space dipole moment of the surface Dirac nodes. This result is a generalization of the notion of characterizing TSMs via multipole moments of the nodal point distribution to the broad class of HOTSMs. In Section~\ref{sec:conclusion} we conclude with a discussion of future directions for this work. 

\section{Model}
\label{sec:model}
In this section we construct a model of a time-reversal symmetric Weyl semimetal in which higher order Weyl nodes are arranged in a quadrupole pattern. We discuss bulk indicators of the topology of this model and the associated bulk, surface, and hinge spectra. Consider the following Bloch Hamiltonian,
\begin{equation}
    \begin{aligned}
        H(\vb{k}) &= \sin(k_x)\sin(k_y)\Gamma_1 + \sin(k_z) \Gamma_2 \\
        &+ \left(m + \cos(k_x) + \beta \cos(k_z)\right)\Gamma_3 \\
        &+ \left(m + \cos(k_y) + \beta \cos(k_z)\right)\Gamma_4 \\
        &+ i \gamma \Gamma_1\Gamma_2,
    \end{aligned}
    \label{eq:H}
\end{equation}
where $\Gamma_i$ is a set of five anti-commuting $4\times4$ matrices. We use the basis $\Gamma_0=\sigma_2\otimes\sigma_0$, $\Gamma_1=\sigma_1\otimes\sigma_1$, $\Gamma_2=\sigma_1\otimes\sigma_2$, $\Gamma_3=\sigma_1\otimes\sigma_3$, and $\Gamma_4=\sigma_3\otimes\sigma_0$, where $\sigma_i$ are the Pauli matrices. This Hamiltonian possesses a range of symmetries: spinless time-reversal symmetry (TRS), $\mathcal{T}=\mathcal{K}\mathbb{I}$, two-fold rotation symmetry about each axis, $C_{2x}=C_{2y}=\Gamma_1\Gamma_2$, $C_{2z}=\mathbb{I}$, mirror symmetry about the $x=y$ and $x=-y$ planes, $M_{1,1}=M_{1,-1}=(\Gamma_3-\Gamma_4)\Gamma_0/\sqrt{2}$, the product of four-fold rotation and reflection along the $z$-axis, $C_{4z}M_z=(\Gamma_3+\Gamma_4)/\sqrt{2}$, and the product of inversion and chiral symmetry, $P\Xi=\Gamma_2$. 

\begin{figure*}
    \centering
    \subfloat[][]{\label{fig:bands_qbc}\includegraphics[width=0.32\textwidth]{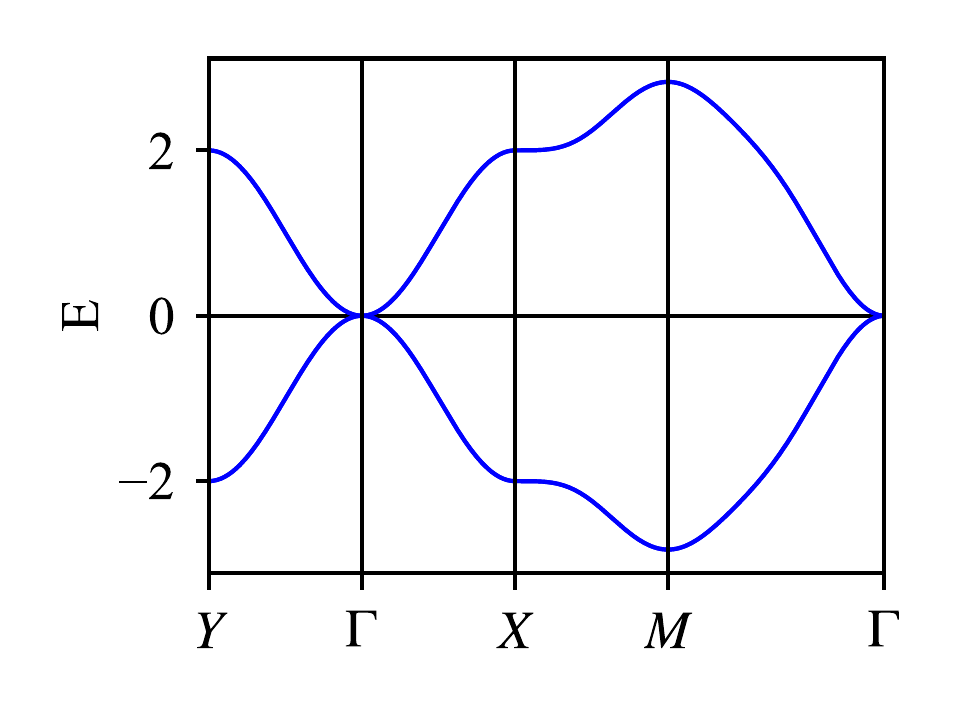}}
    \subfloat[][]{\label{fig:bands_weyl}\includegraphics[width=0.32\textwidth]{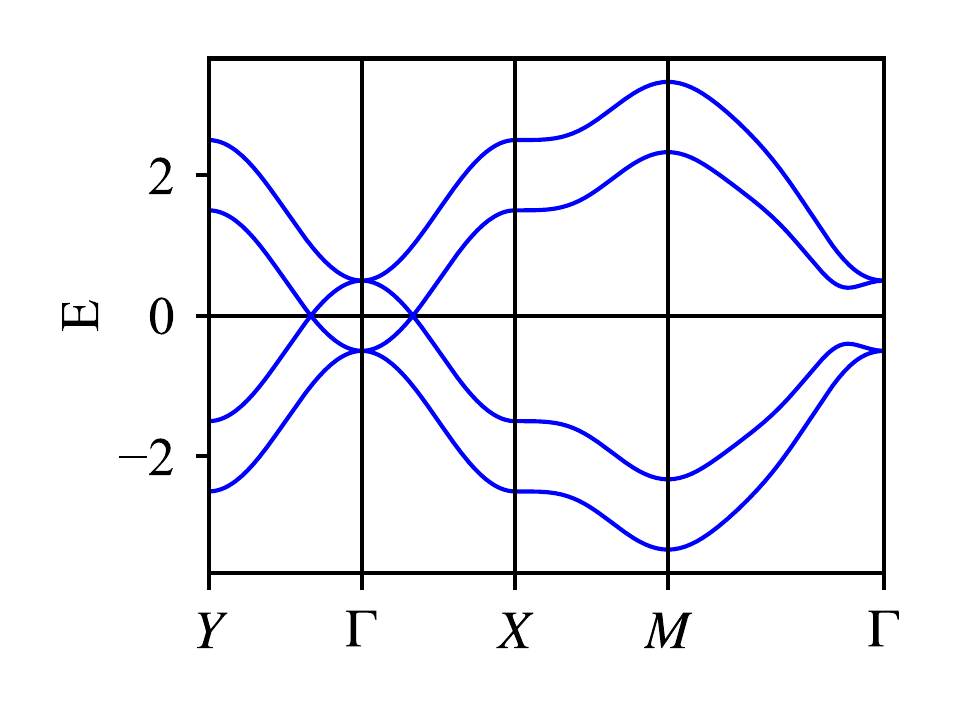}}
    \\
    \subfloat[][]{\label{fig:chern_slices}\includegraphics[width=0.32\textwidth]{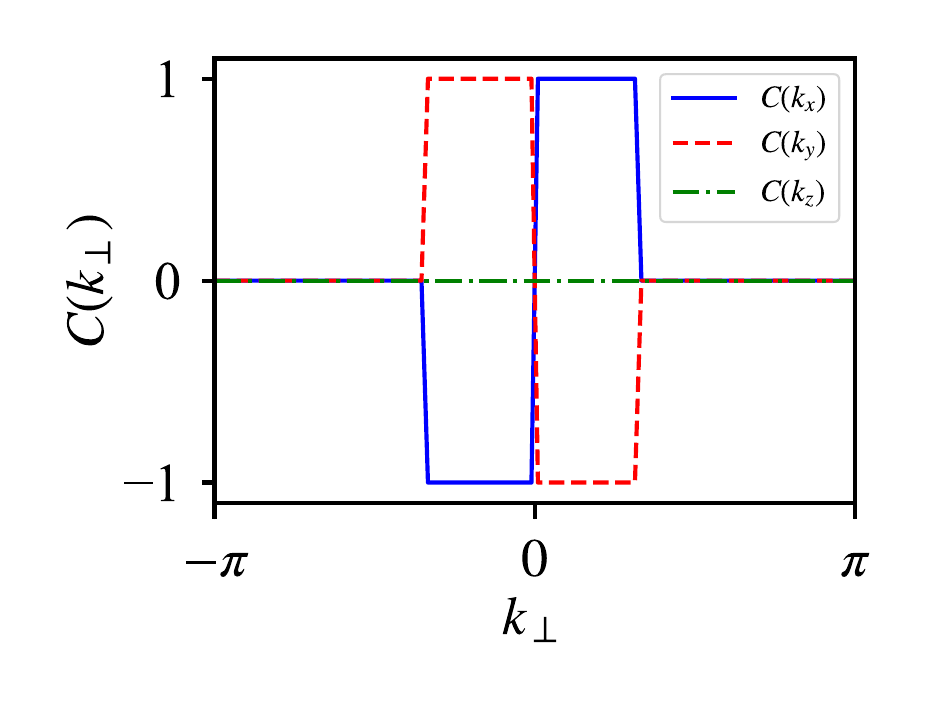}}
    \hfill
    \subfloat[][]{\label{fig:nested_wilson_loops}\includegraphics[width=0.33\textwidth]{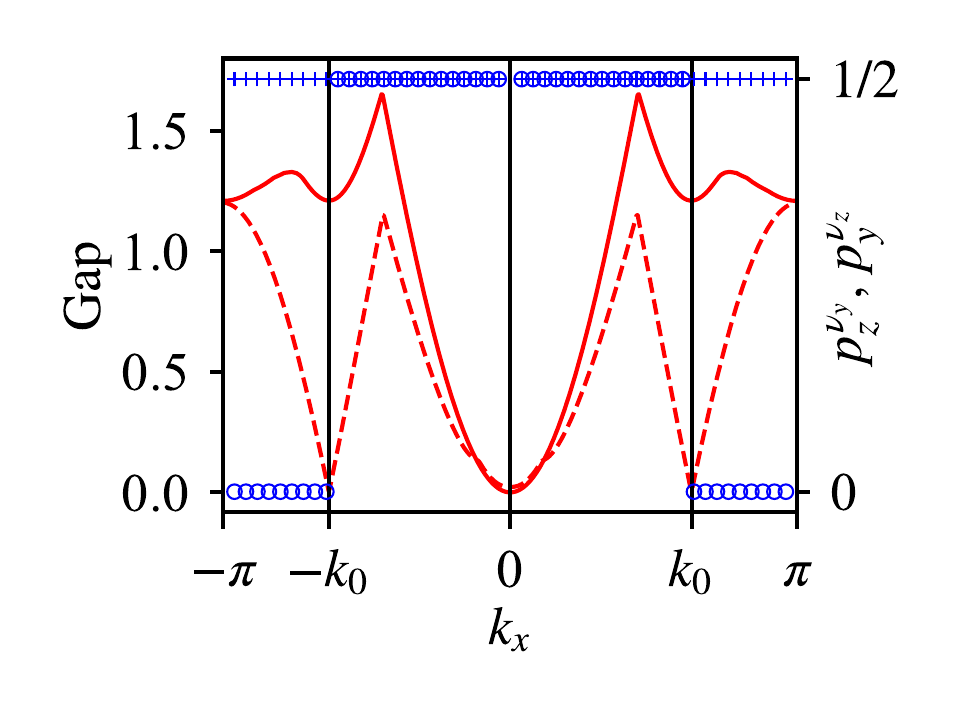}}
    \hfill
    \subfloat[][]{\label{fig:kx_gamma_phase_diagram}\includegraphics[width=0.33\textwidth]{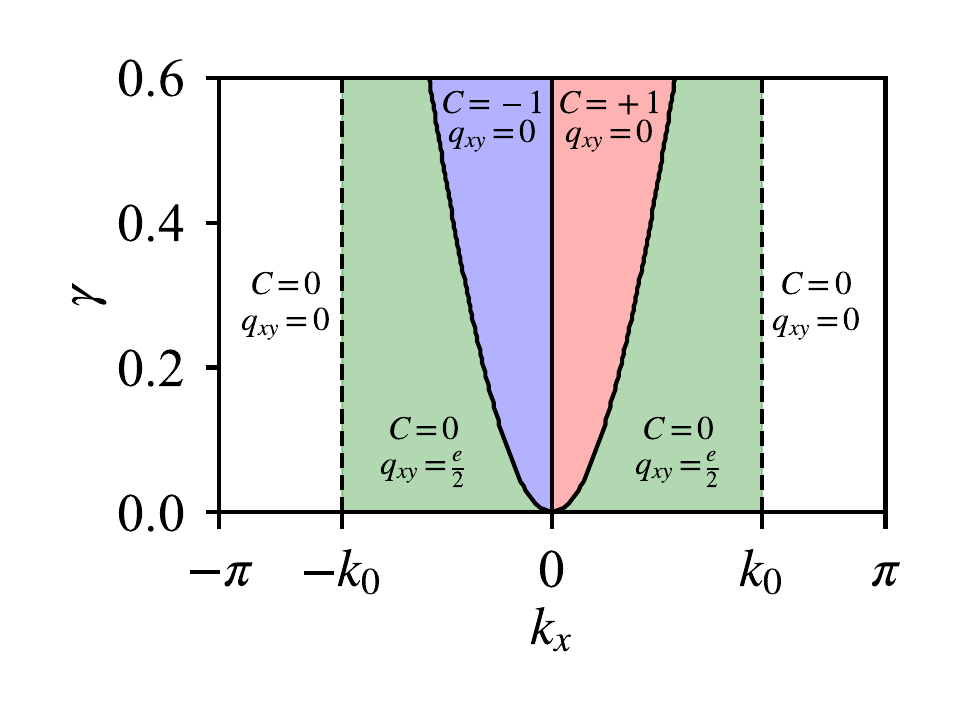}}
    \caption[]{(a) The band structure of $H(\vb{k})$ along high-symmetry lines in the $k_z=0$ plane with $m=-0.3$, $\beta=-0.7$, and $\gamma=0$. The band structure possesses a QBC at $\Gamma$ and is otherwise gapped. (b) The band structure of $H(\vb{k})$ with $m=-0.3$, $\beta=-0.7$, and $\gamma=0.5$. The finite value of $\gamma$ splits the QBC into four Weyl nodes, two on the $k_x$ axis and two on the $k_y$ axis. (c) The Chern number of $H(\vb{k}; k_x)$ (solid blue), $H(\vb{k}; k_y)$ (dashed red), and $H(\vb{k}; k_z)$ (dot-dashed green) as functions of the perpendicular momentum with $=m-0.3$, $\beta=-0.7$, and $\gamma=0.5$. The changes in the Chern number as the perpendicular momenta are tuned through Weyl nodes indicates that the nodes along $k_x$ and $k_y$ are of negative and positive chirality, respectively. (d) The nested Wilson loops $p_z^{v_y}$ (blue crosses) and $p_y^{v_z}$ (blue circles), bulk gap (solid red line), and surface gap (dashed red line) of $H(\vb{k}; k_x)$ with $m=-0.3$, $\beta=-0.7$, and $\gamma=0.0$. (e) The finite-$\gamma$ phase diagram of $H(\vb{k}; k_x)$ with $m=-0.3$, and $\beta=-0.7$. The solid and dashed black lines indicate bulk and surface gap closings of $H(\vb{k}; k_x)$, respectively. The light green region is adiabatically connected to the $\gamma=0$ QI phase and therefore has $C=0$ and $q_{xy} = e/2$. The red and blue regions are Chern insulator phases with $C=\pm1$, and the white regions are trivial.}
    \label{fig:bulk_bands}
\end{figure*}

We first consider the bulk energy spectrum of $H(\vb{k})$ in the special case $m+\beta=-1$ and $\gamma=0$, for which a quadratic band crossing (QBC) appears at $\Gamma$, as shown in Fig.~\ref{fig:bands_qbc}. While $m$ and $\beta$ can be tuned to generate QBCs at other high-symmetry points of the BZ, we only consider parameter ranges that place the QBC at $\Gamma$. Departing from this starting point by tuning $\gamma$ away from zero splits the QBC into four  Weyl nodes that move apart along the $k_x$ and $k_y$ axes. We show the finite-$\gamma$ spectrum in Fig.~\ref{fig:bands_weyl} which clearly depicts the Weyl nodes on the $\Gamma X$ and $\Gamma Y$ lines.

To identify the bulk topology, we recall that Weyl nodes act as quantized sources of Berry curvature. As such, the Chern number of any surface in momentum space that encloses a single Weyl node is $C=\pm1$, where the sign is determined by the chirality $\chi$ of the node. Consequently, we can foliate the Brillouin zone into families of fixed momentum planes, and planes that are separated by a Weyl node must have Chern numbers differing by $\chi.$  This planar family picture is very convenient and we denote the Hamiltonian restricted to two-dimensional momentum planes normal to the $k_i$ axis as $H(\vb{k};k_i)$. In Fig.~\ref{fig:chern_slices} we plot the Chern numbers of $H(\vb{k}, k_i)$ for $i=x, y, z$ as functions of $k_i$ with $m=-0.3$, $\beta=-0.7$ and $\gamma=0.5$. The discrete jumps in Chern number at the Weyl nodes indicate that the chiralities of the nodes on the $k_x$ and $k_y$ axes are negative and positive, respectively.

\begin{figure*}
    \centering
    \subfloat[][]{\label{fig:001_surf_bands}\includegraphics[width=0.32\textwidth]{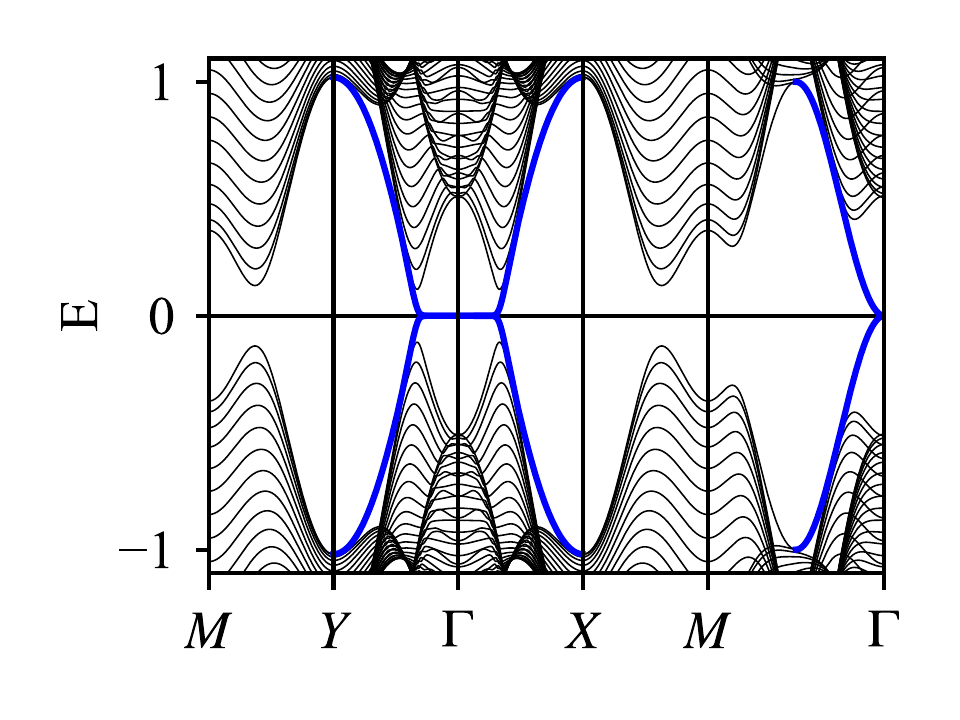}}
    \hfill
    \subfloat[][]{\label{fig:100_surf_bands}\includegraphics[width=0.32\textwidth]{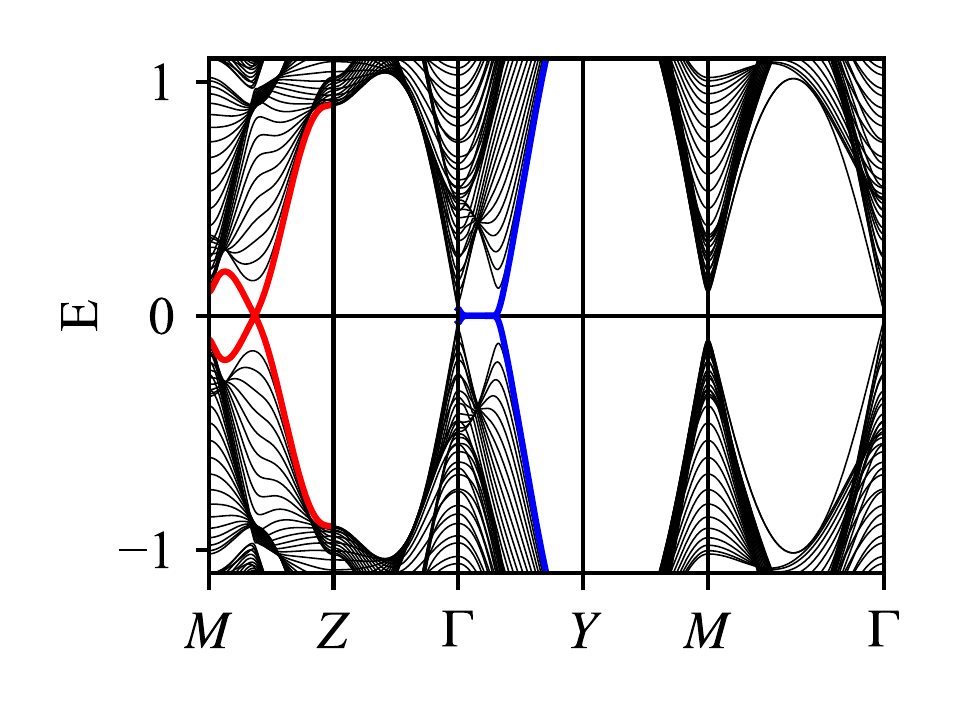}}
    \hfill
    \subfloat[][]{\label{fig:100_hinge_bands}
    \includegraphics[width=0.32\textwidth]{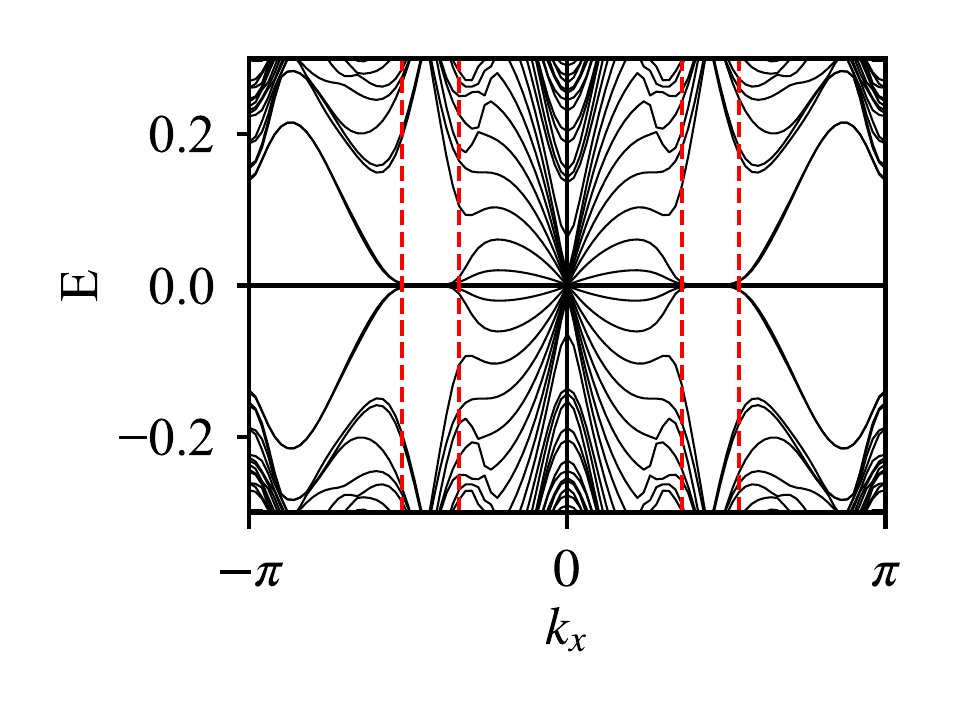}}
    \caption[]{The (a) $z$- and (b) $x$-normal surface band structures of Eq.~\ref{eq:H} along high-symmetry lines with $m=-0.3$, $\beta=-0.7$ and $\gamma=0.5$, using 30 lattice sites in the open direction. The $z$-normal surface has a cross of Fermi arcs connecting the projections of the Weyl nodes on both the $k_x$ and $k_y$ axes. The $x$-normal surface possesses Fermi arcs between the projections of the Weyl nodes on the $\Gamma-Y$ line and a pair of Dirac cones on the BZ boundary. Bands containing Fermi arcs are drawn in blue and the surface Dirac cones are indicated with red. The spectrum of the $y$-normal surface is identical to the $x$-normal surface. (c) The spectrum of $H(\vb{k})$ with open boundary conditions along the $y$- and $z$-directions, 25 lattice sites along each open direction, $m=-0.3$, $\beta=-0.7$, and $\gamma=0.5$. The zero-energy modes arise from the quadrupole phases of $H(\vb{k};k_x)$ and are localized to the hinges. The dashed red lines indicate the bounds of the zero-energy hinge modes. The hinge spectrum along the $y$-direction is identical.}
    \label{fig:surf_bands}
\end{figure*}

The fixed-momentum planes having non-vanishing Chern number generate chiral edge modes along open boundaries. The collection of these edge states comprise the surface Fermi arcs that connect projections of the Weyl nodes in the surface BZ. In Fig.~\ref{fig:001_surf_bands} we plot the surface spectrum of $H(\vb{k})$ with open boundary conditions along the $z$-direction. At zero energy there are a pair of intersecting Fermi arcs, which we depict in blue, on the surface normal to the $z$-direction, with one nodal arc on the $k_x$-axis and another arc on the $k_y$-axis. At energies above or below $E=0$ the Fermi arcs form portions of a hyperbola that originate at the positive chirality nodes, nearly meet at the origin, and then turn in opposite directions to eventually terminate at the negative chirality nodes. Indeed, the dispersion around $\Gamma$  is that of a saddle point $E= k_x k_y$, hence this model is another realization of a surface rank-2 chiral fermion~\cite{dubinkinHigherRankChiral2021}. For comparison, in Fig.  \ref{fig:100_surf_bands} we plot the surface spectrum with open boundaries in the $x$-direction with $m=-0.3$, $\beta=-0.7$, and $\gamma=0.5.$ We find that the Fermi arcs that appear on the $x$-normal surface originate at $\Gamma$ and terminate at the positive- and negative-momentum projections of the Weyl nodes on the $k_y$-axis. On a $y$-normal surface the relative chirality of the nodes switches, but the Fermi arcs  are identical because of the mirror and rotational symmetries of $H(\vb{k})$.

We can characterize arrangements of Weyl nodes by calculating the momentum-space multipole moments of the nodes weighted by the node chiralities. In particular, we define the Weyl dipole $P_i$ and Weyl quadrupole $Q_{ij}$ moments as
\begin{equation}
    P_a = \sum_n \chi^n k^n_a,\quad Q_{ab} = \sum_n \chi^n k^n_ak^n_b,
\end{equation}
where $n$ indexes the nodes. The Weyl dipole moment of $H(\vb{k})$, which is proportional to the anomalous Hall conductivity,  vanishes as required by TRS. In contrast, we find that the diagonal quadrupole moments $Q_{xx}$ and $Q_{yy}$ are non-vanishing, and the mirror symmetries along the $x=y$ and $x=-y$ axes require them to have the same magnitude and opposite sign.  We consider these moments because, as mentioned above, the dipole moment is directly related to the anomalous Hall coefficient, and recent works have shown that the quadrupole moment characterizes mixed crystalline-electromagnetic responses, e.g., screw dislocations bind electric charge, and magnetic flux binds crystal momentum~\cite{dubinkinHigherRankChiral2021, gioia_unquantized_2021}. Below we show that the Weyl nodes in our model are, in fact, higher order Weyl nodes, and investigate the mixed crystalline-electromagnetic responses that arise from quadrupole arrangements of higher order Weyl nodes.

We have mentioned that first order Weyl nodes represent a transition (as a function of momentum) between insulator phases on planes of the foliated BZ where the Chern number differs by the Weyl chirality. In contrast, higher order Weyl nodes separate insulators that differ by both a Chern number and some type of 2D higher order topology. In our case the higher order topology is that of a quadrupole insulator (QI)~\cite{benalcazarElectricMultipoleMoments2017, benalcazarQuantizedElectricMultipole2017, peng_boundary_2017, song_dimensional_2017, schindler_higher_2018, langbehn_reflection_2017, khalaf_symmetry_2018, khalaf_shift_2019, benalcazar_quantization_2019, roy_higher_2019}. Depending on the symmetry, such QI phases can be either bulk obstructed or boundary obstructed~\cite{khalaf_boundary_2021, ezawa_edge_2020, asaga_boundary_2020, benalcazar_chiral_2022, tiwari_chiral_2020}, and they are characterized by a quantized bulk electric quadrupole moment $q_{xy}=e/2$ and a quantized, vanishing bulk charge polarization. The bulk electric quadrupole moment is defined as $q_{xy} = p^{\partial}_x + p^{\partial}_y - Q_{corner} \mod 1$, where $p^{\partial}_x$ and $p^{\partial}_y$ are the electric polarizations on $\hat{y}$- and $\hat{x}$-normal surfaces, respectively, and $Q_{corner}$ is the charge localized on a corner where two such surfaces meet. One typical manifestation of a bulk electric quadrupole moment $q_{xy}$ is a set of corner charges in systems with open boundary conditions in both the $x$- and $y$-directions. For our model these corner charges are accompanied by a set of four mid gap corner modes, the occupation of which determines the pattern of signs of the corner charges.

To show that the Weyl nodes in our model are higher order we need a procedure to diagnose the QI topology. One approach is to study the pair of Berry phases $(p_x^{\nu_y}, p_y^{\nu_x})$ of the hybrid Wannier bands $v_y(k_x)$ and $v_x(k_y)$~\cite{benalcazarElectricMultipoleMoments2017, benalcazarQuantizedElectricMultipole2017}. These Berry phases, which are referred to as nested Wilson loops, indicate the QI phase with non-vanishing $q_{xy}$ when they are both non-trivial, i.e., when  $(p_x^{\nu_y}, p_y^{\nu_x})=(1/2, 1/2)$. The symmetry restrictions required to quantize the nested Wilson loops are more stringent than those required to enforce a non-vanishing, quantized quadrupole moment, so this approach can be applied only in a reduced parameter region of our model. Typically, a pair of mirror symmetries is needed to quantize the nested Wilson loops, but our putative QI insulator Hamiltonians $H(\vb{k},k_x)$ and $H(\vb{k},k_y)$ instead possess pairs of mirror \emph{times time-reversal} symmetries. These symmetries are represented by $M_{x/y}\mathcal{T}=\mathbb{I}_{4\times4}$ and $M_z\mathcal{T}=\Gamma_1\Gamma_2$, and descend from the the $C_{2x/y}$, $C_{2z}$, and $\mathcal{T}$ symmetries of $H(\vb{k})$. These mirror times time-reversal symmetries quantize the bulk quadrupole moment but do not quantize the nested Wilson loops. 

We can make progress by noting that these symmetries are elevated to conventional mirror symmetries in the limit $\gamma=0.$ Hence the $\gamma=0$ limit permits the computation of the bulk quadrupole moment via the nested Wilson loops. We present the results of this computation in Fig.~\ref{fig:nested_wilson_loops} where we plot the bulk gap, surface gap, and nested Wilson loops of $H(\vb{k}, k_x)$ with $m=-0.3$, $\beta=-0.7$, and $\gamma=0$ as a function of $k_x$. We find that the bulk gap closes at $k_x=0$, corresponding to the QBC at $\Gamma.$ Interestingly, the surface gap closes at a pair of momenta $k_x=\pm k_0$, far away from the location of the bulk gap closing. For $0 < |k_x| < |k_0|$, both nested Wilson loops are quantized to $1/2$, confirming the presence of a non-trivial QI phase for each fixed-$k_x$ plane in this interval. One of the two nested Wilson loops changes values at the surface gap closing at $|k_x|=k_0$, leaving the region $|k_x| > k_0$ with only a single non-trivial nested Wilson loop, indicating a phase with vanishing quadrupole moment for all fixed-$k_x$ planes in this interval.

While we can only calculate the quantized nested Wilson loops for $\gamma=0,$ we can go beyond the $\gamma \neq 0$ case by using an adiabatic argument. As long as the crystal symmetries that quantize $q_{xy}$ and the $x,y$ components of the (bulk) polarization are maintained, the bulk quadrupole moment can change only at bulk or surface gap closing points. Thus, knowing the results for $\gamma=0,$ we can determine the bulk quadrupole moment at finite $\gamma$ via a straightforward adiabatic argument. At any momentum $k_x$ for which $H(\vb{k}, k_x)$ realizes the $\gamma=0$ QI phase, the Hamiltonian will remain in the QI phase at finite $\gamma$ as long as there are no intervening bulk or surface gap closings, and the quantizing symmetry is maintained. We plot the locations of the bulk and surface gap closings of $H(\vb{k}, k_x)$ in Fig.~\ref{fig:kx_gamma_phase_diagram} as a function of $k_x$ and $\gamma$ with $m=-0.3$ and $\beta=-0.7$. The splitting of the QBC into Weyl nodes nucleates a pair of $C=-1$ and $C=+1$ Chern insulator phases on opposite sides of $k_x=0$, indicated in blue and red, respectively. The locations of the surface gap closings do not depend on $\gamma$, so the QI remains intact for $k_{\mathrm{Weyl}} < |k_x| < k_0$, where $k_{\mathrm{Weyl}}$ is the location of the Weyl node on the $k_x$ axis. Similar results obtain when we consider the 2D, fixed-momentum phases as a function of $k_y$ instead of $k_x.$ This confirms that the Weyl nodes in this system separate Chern insulator phases from QI phases and are higher order Weyl nodes.

The surface gap closings that bound the QI phases of $H(\vb{k}, k_x)$ appear as a pair of surface Dirac cones at opposite values of $k_x$ on the $k_z=\pi$ boundary of the $y$-normal surface BZ. An analogous pair of surface Dirac cones appears on $x$-normal surface BZs owing to the rotation and mirror symmetries of $H(\vb{k})$. We plot the $x$-normal surface spectrum with $m=-0.3$, $\beta=-0.7$, and $\gamma=-0.5$ in Fig.~\ref{fig:100_surf_bands}, in which the surface Dirac cone at positive $k_y$ is visible and depicted in red. With open boundary conditions along both the $y$- and $z$-directions, the hinge spectrum of $H(\vb{k})$, shown in Fig.~\ref{fig:100_hinge_bands}, exhibits a pair of mid-gap flat bands in the hinge BZ spanning between the projections of the bulk Weyl nodes and the surface Dirac nodes. These mid-gap hinge arcs originate from the mid-gap corner modes of the QI phase. We find identical results for hinges parallel to $\hat{y}$ as ensured by the mirror and rotation symmetries of $H(\vb{k})$. In the next section we study the mixed crystalline-electromagnetic responses that arise from such quadrupole arrangements of higher order Weyl nodes, with $H(\vb{k})$ serving as an explicit realization. Additionally, in Sec.~\ref{sec:momentum_quadrupole_moment} we study some further consequences of the mid-gap hinge states.

\section{Mixed Charge-Momentum Responses}
\label{sec:mixed_responses}

It was recently shown that semimetals hosting a quadrupole configuration of Weyl nodes exhibit a mixed charge-momentum response that binds crystal momentum to magnetic flux and electric charge to screw dislocations~\cite{dubinkinHigherRankChiral2021, gioia_unquantized_2021}. Here we confirm that the Hamiltonian Eq.~\eqref{eq:H} also exhibits this response. Furthermore, we show that the higher order nature of our model's Weyl nodes leads to an additional \emph{surface} mixed charge-momentum response. This surface response manifests as crystal momentum bound to magnetic flux and electric charge bound to dislocations.

The mixed charge-momentum response of topological semimetals hosting a Weyl quadrupole is captured by the effective action 
\begin{equation}
    S[A,\mathfrak{e}] = -\frac{e}{8\pi^2}\int d^4x \, \epsilon^{\mu\nu\rho\sigma}Q_{a b} \mathfrak{e}^a_\mu A_\nu\partial_\rho \mathfrak{e}^b_\sigma,
    \label{eq:bulk_effective_action}
\end{equation}
where $Q_{ab}$ is the quadrupole moment of the Weyl nodes, $A_\mu$ is the electromagnetic gauge field, and $\mathfrak{e}_\mu$ are the translation gauge fields~\cite{PhysRevLett.107.075502,thorngrenGauging2018, nissinenElasticity2019, nissinenTopologicalPolarizationDual2021, song_electric_2021, manjunathCrystalline2021}. For our model we simplify this action by noting that only the diagonal elements of the quadrupole moment of $Q_{ab}$ are non-vanishing $Q_{xx} = -Q_{yy} \equiv \bar{Q}$. 

One response encoded by this action is the binding of momentum density to magnetic flux that points along the $x$- or $y$-directions,
\begin{equation}
    \mathcal{J}^0_a = \frac{e\bar{Q}}{8\pi^2} B_a\left(\delta_{ax} - \delta_{ay}\right),
\end{equation}
where the bound momentum points along the magnetic field and the momentum density of the electrons is defined as
\begin{equation}
    \mathcal{J}^0_a = \frac{1}{e}\int \frac{d^3\vb{k}}{(2\pi)^3}\, k_a j^0(\vb{k}).
\end{equation}
There is also a conjugate response wherein charge is bound to screw dislocations that have Burgers vectors in the $xy$-plane,
\begin{equation}
    j^0 = \frac{e\bar{Q}}{8\pi^2} \left(\mathcal{B}^x_x - \mathcal{B}^y_y\right).
\end{equation}
In these two response equations $B_a$ are the components of the magnetic field, $\mathcal{B}_i^i = \epsilon^{ijk}\partial_j\mathfrak{e}_k^i$ is the torsional magnetic field induced by a screw dislocation along the $i$-axis, and we set the diagonal components of the translation gauge field equal to their background values, i.e., $\mathfrak{e}^x_x=\mathfrak{e}^y_y=\mathfrak{e}^z_z=1,$ which encode the existence of the discrete Bravais lattice. 

This mixed-charge momentum response can be straightforwardly understood as a consequence of the arrangement of non-trivial Chern insulator phases on  planes in the foliated BZ. For simplicity, let us first consider planes normal to $k_x$ and denote the locations of the Weyl nodes away from $k_x=0$ on the $k_x$-axis as $\pm k_0$. As shown in Fig.~\ref{fig:chern_slices}, the Chern number of $H(\vb{k};k_x)$ is $C=-1$ for $-k_0 < k_x < 0$, $C=1$ for $0 < k_x < k_0$, and $C=0$ elsewhere. Consider inserting a magnetic flux $\Phi$ in the $yz$-plane. Let us assume that this flux line preserves translation symmetry along $\hat{x}.$ Then the net response of the system to the magnetic flux is the response of $H(\vb{k},k_x)$ summed over $k_x$. The trivial phases of $H(\vb{k},k_x)$ are inert to the flux, but the Chern insulator phases bind charge $q=C\Phi/\Phi_0$ to the flux, where $\Phi_0$ is the quantum of magnetic flux~\cite{stredaTheoryQuantisedHall1982}. The charge density bound to the flux by the $C=1$ and $C=-1$ phases of $H(\vb{k},k_x)$ are opposite, so no net charge is accumulated. However, the crystal momentum-per-length bound to the flux is non-vanishing:
\begin{equation}
    \int dydz \, \mathcal{J}_x^0 = \frac{e\bar{Q}}{8\pi^2}\Phi.
\end{equation}

The dual response of charge bound to a screw dislocation along the $\hat{x}$-direction can be understood through similar reasoning. As with the Aharonov-Bohm effect for electrons near a magnetic flux line, electrons encircling a screw dislocation acquire a phase. In the magnetic flux case the phase is proportional to a product of the charge and flux, $\varphi \propto  e\Phi.$ In the translation flux case, the phase is the dot product of the crystal momentum of the electron (translation charge) and the Burgers vector of the dislocation (translation flux), $\varphi = \vb{k} \cdot \vb{b}$, where $\vb{b}=(b_x, 0, 0)$ in this case. Since the phase acquired upon encircling the screw dislocation is proportional to $k_x$, the $C=1$ and $C=-1$ phases of $H(\vb{k},k_x)$ bind \emph{equal} charge (in both sign and magnitude) to the defect, yielding no bound crystal momentum density. However, there is a non-vanishing bound charge-per-length:
\begin{equation}
    \int dydz\, j_0(\vb{r}) = \frac{eQ_{xx}}{8\pi^2}b_x.
\end{equation}

The response to threading magnetic flux or screw dislocations along other directions can be interpreted similarly. That is, one can determine the arrangement of the Chern insulator phases perpendicular to the chosen direction $\hat{n}$ by projecting the Weyl nodes onto that axis in momentum space. Then one can apply the flux insertion method above to determine the response. As an additional example, this model  has the interesting characteristic that for $\hat{n} = \hat{x} \pm \hat{y}$ and $\hat{n} = \hat{z}$, the response is zero because the Weyl nodes project onto the given axes in opposite-chirality pairs, yielding $C=0$ for all momenta.

Since our model $H(\vb{k})$ has Weyl nodes arranged in a quadrupolar pattern we expect to find it has the responses encoded by Eq. \ref{eq:bulk_effective_action}. Here we verify that $H(\vb{k})$ exhibits the mixed charge-momentum response described above by numerically calculating both the electric charge density bound to screw dislocations and the momentum density bound to magnetic fluxes. We consider a system with periodic boundary conditions in all directions and choose a configuration to preserve translation symmetry along $\hat{x}$, which is necessary to permit calculation of the crystal momentum density along $\hat{x}$. As such, we treat the $x$-direction in momentum space with $N_k=40$, and use a lattice of dimension $N_y\times N_z = 40\times 40$ in the $y$-and $z$-directions. We insert oppositely-signed flux lines, either electromagnetic or translational, along $\hat{x}$ at sites $(y,z)=(20, 10)$ and $(20, 30)$. 

To generate the fluxes we include the magnetic flux $\Phi$ via a Peierls phase, i.e., multiplying all hopping terms that cross the line connecting the two flux lines by the phase $\exp\left(2\pi i\Phi/\Phi_0\right)$. Because the translation gauge fields couple to momentum rather than charge, the translational magnetic field of a screw dislocation is accounted for by modifying the Peierls phases used for the magnetic flux to be a product of the crystal momentum along $\hat{x}$ and the translational flux of the dislocation, $k_x\Phi^T$~\cite{PhysRevLett.107.075502,PhysRevD.88.025040,PhysRevD.90.105004}. The modified Peierls phase captures the phase acquired by an electron having crystal momentum $k_x$ encircling the screw dislocation and translating by $\Phi^T$ sites in the $\hat{x}$-direction. 

\begin{figure*}
    \centering
    \subfloat[][]{\label{fig:magnetic_bulk}\includegraphics[width=0.32\linewidth, valign=c]{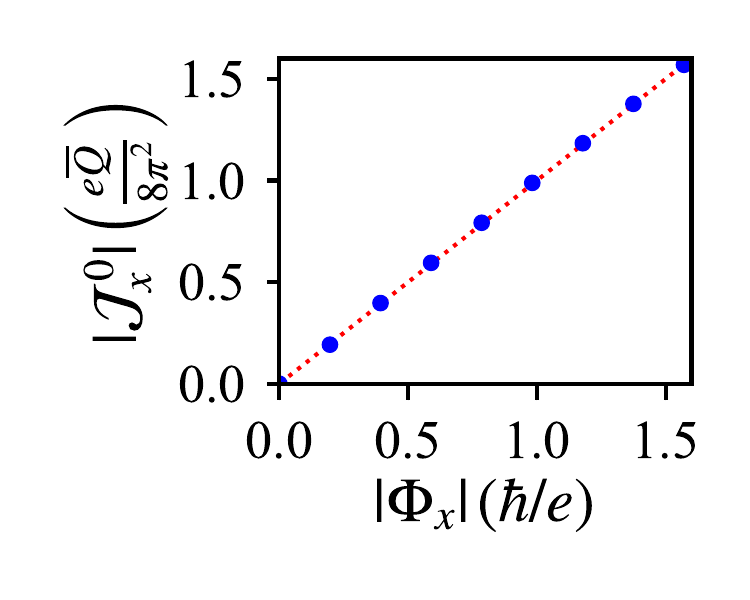}}
    \hspace*{0.01\linewidth}
    \subfloat[][]{\label{fig:magnetic_surface}\includegraphics[width=0.32\textwidth, valign=c]{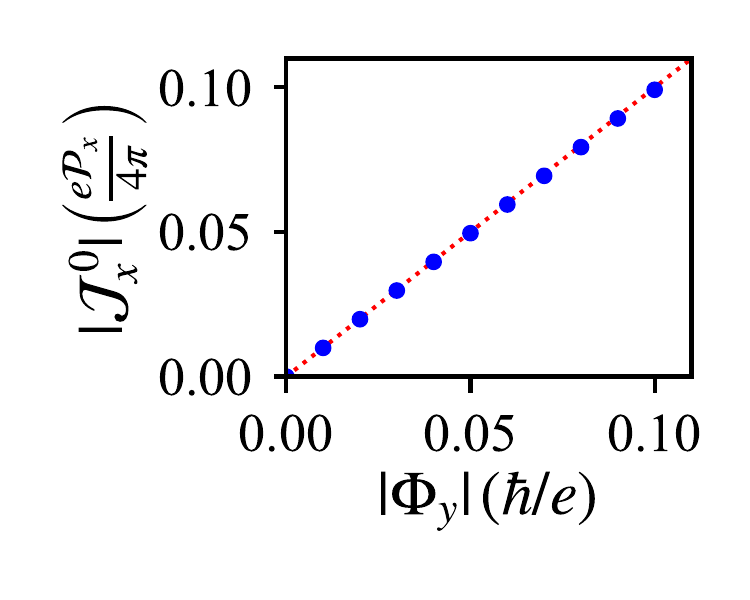}}
    \hspace*{0.025\linewidth}
    \subfloat[][]{\label{fig:field_geometry_3d}\includegraphics[width=0.3\textwidth, valign=c]{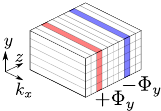}}
    \\
    \subfloat[][]{\label{fig:torsional_bulk}\includegraphics[width=0.32\linewidth, valign=c]{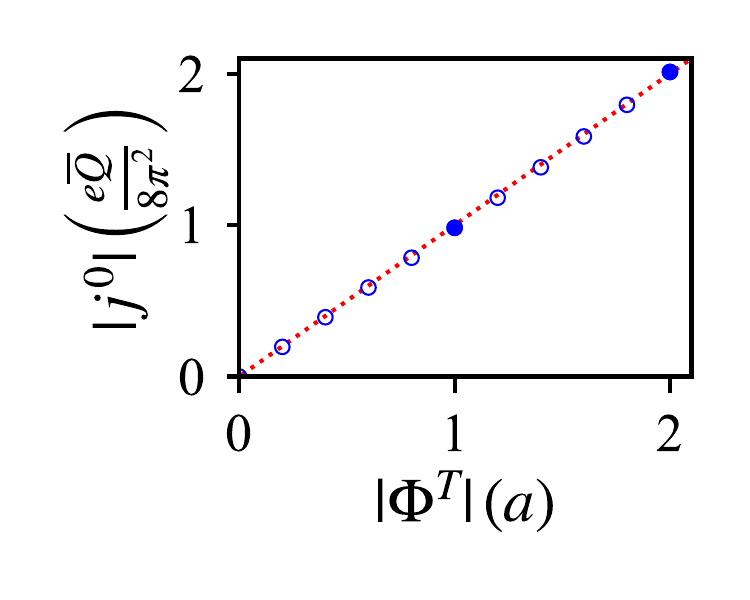}}
    \hspace*{0.01\linewidth}
    \subfloat[][]{\label{fig:dislocation_surface}\includegraphics[width=0.32\textwidth, valign=c]{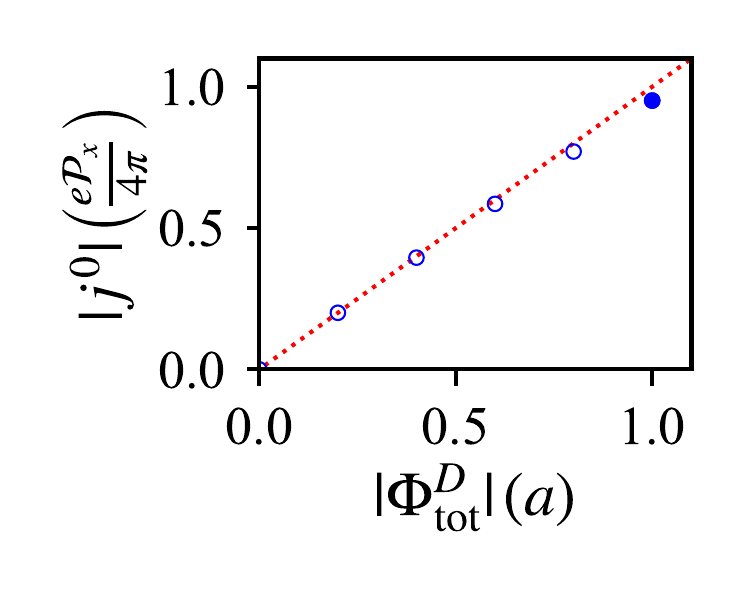}}
    \hspace*{0.025\linewidth}
    \subfloat[][]{\label{fig:grain_boundary}\includegraphics[width=0.25\textwidth, valign=c]{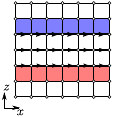}}
    \caption[]{(a) The crystal momentum bound to magnetic flux by Eq.~\eqref{eq:H} on a lattice of dimension $N_y\times N_z = 40\times 40$, and $N_{k_x}=40$. (b) The surface crystal momentum density $\mathcal{J}^0_x$ bound to magnetic flux along the $y$-direction for a system size of $N_y\times N_z=30\times30$ and $N_{k_x} = 40$. (c) The flux geometry we use to calculate the momentum and charge density response on $x$- and $y$-normal surfaces. Red (blue) coloration indicates either magnetic or dislocation flux pointing along the $+y$-direction ($-y$-direction). (d) The electric charge bound to a screw dislocation with $N_y\times N_z = 40\times 40$, and $N_{k_x}=40$. Empty and filled circles indicate fractional and integer torsional fluxes. (e) The surface electric charge density bound to dislocation flux for a system size of $N_y\times N_z=40\times 40$ and $N_{k_x}=100$. Here $\Phi^D_{\mathrm{tot}}$ is the total dislocation flux integrated along the $x$-direction, corresponding to the difference in system size along the $x$-direction between the strained and unstrained regions in units of the unstrained lattice constant. Empty and filled circles indicate fractional and integer dislocation fluxes. (f) The dislocation flux geometry used to calculate the surface charge response. The red (blue) plaquettes correspond to positive (negative) dislocation flux and the black arrows indicate the hoppings that acquire a momentum-dependent Peierls phase due to the strain of the lattice. All results presented here are calculated using the parameters $m = \beta = -0.5$ and $\gamma = 0.5$. The red dotted lines in (a), (b), (d), and (e) each have a slope of one and indicate the analytic result.}
    \label{fig:bulk_response}
\end{figure*}

Using this setup, we calculate the momentum density bound to magnetic flux as a function of the flux as shown in Fig.~\ref{fig:magnetic_bulk}. Similarly, in Fig.~\ref{fig:torsional_bulk} we plot the charge bound to a screw dislocation as a function of the translational magnetic flux. Both plots demonstrate the expected linear relationship between charge/momentum and flux with slope $e\bar{Q}/8\pi^2$, corroborating that the Hamiltonian $H(\vb{k})$ possesses the response predicted by the effective action in Eq.~\eqref{eq:bulk_effective_action}. Let us comment about the data points represented by open circles in Fig.~\ref{fig:torsional_bulk}. In order to be commensurate with the lattice, the torsional flux $\Phi^T$ should take integer values and is equivalent to the Burgers vector of the screw dislocation. The open circle data points are non-integer translation fluxes that can be inserted into our momentum-dependent Peierls factors, but the interpretation in terms of an elastic lattice defect is less clear. 

Now that we have confirmed the expected bulk responses we can move on to identify the surface responses. Indeed, as a consequence of our Weyl nodes being higher order, we expect that even regions of the surface BZ that do not harbor gapless surface states may contribute to surface responses. Indeed, since the $\hat{x}$- and $\hat{y}$-normal surfaces of the model Hamiltonian host a pair of Dirac nodes we expect to find a 2D surface response analogous to a 2D Dirac semimetal. As such, these surfaces possess a mixed charge-momentum response similar to that of the bulk described by the effective action~\cite{ramamurthy_patterns_2015}:
\begin{equation}
    S[A,\mathfrak{e}] = \frac{e\mathcal{P}_a}{4\pi}\int d^4x \epsilon^{\mu\nu\rho}\,\mathfrak{e}^a_\mu \partial_\nu A_\rho.
    \label{eq:surface_effective_action}
\end{equation}
Here the response coefficient is the Berry curvature dipole moment, given by~\cite{ramamurthy_patterns_2015, song_electric_2021, dubinkin_in_prep}
\begin{equation}
    \mathcal{P}_a = \frac{1}{\pi}\int_{\mathrm{BZ}}d^2\vb{k} k_a \mathcal{F}(\vb{k}),
\end{equation}
where $\mathcal{F}$ is the Berry curvature and the integration is restricted to the surface BZ. When the surface has time-reversal and inversion symmetry, this action implies that the system has a charge polarization  $p^a=\frac{e}{4\pi}\epsilon^{ab}\mathcal{P}_b$\cite{ramamurthy_patterns_2015} (which resides on the surface of our 3D system).

To illustrate a particular response let us focus on the response of the $y$-normal surface, for which  $\mathcal{P}_x\neq 0$ and $\mathcal{P}_z=0$ (the $x$-normal surface has an analogous response by symmetry). The mixed charge-momentum response captured by the effective action Eq.~\eqref{eq:surface_effective_action} binds momentum density to magnetic flux,
\begin{equation}
    \mathcal{J}_x^0 = -\frac{e}{4\pi}\mathcal{P}_x B_y,
\end{equation}
and binds electric charge to dislocations,
\begin{equation}
    j^0 = -\frac{e}{4\pi}\mathcal{P}_x\left(\partial_x\mathfrak{e}^x_z - \partial_z\mathfrak{e}^x_x\right).
\end{equation}
Here we verify that the surfaces of our model have these responses via direct numerical calculation. We consider a system with open boundary conditions in the $\hat{y}$-direction and periodic boundary conditions in the $\hat{x}$- and $\hat{z}$-directions. We treat the $\hat{x}$-direction in momentum space and calculate the momentum density bound to magnetic flux (using $N_k=40$ momentum points), and the charge bound to dislocations (using $N_k=100$ momentum points). The other two directions we leave in position space and use a lattice of dimension $N_y\times N_z=30\times30$ and $N_y\times N_z=40\times40$ for each of the calculations respectively. To avoid difficulties arising from the divergent Berry curvature distribution of Dirac nodes, we also include an inversion-breaking perturbation $H'=-\nu\frac{i}{2}\Gamma_2\Gamma_3$ with $\nu=0.5$.

To calculate the $k_x$ momentum density on a $\hat{y}$-normal surface it is necessary to maintain translation symmetry along $\hat{x}.$ To do so we introduce the magnetic field via two strips of magnetic flux lines extending in the $x$-direction, each with opposite field orientations $\pm B_y\hat{y}$. For the $N_y\times N_z=40\times40$ lattice these strips are located at $z_1=10$ and $z_2=30$, while for the $N_y\times N_z=30\times30$ lattice they are located at $z_1=8$ and $z_2=23$. The geometry of this magnetic field configuration is depicted in Fig.~\ref{fig:field_geometry_3d}. We include dislocation flux, $\Phi_D$, in an analogous translation symmetry-preserving manner by using the generalized, momentum-dependent Peierls factors mentioned above for the bulk response. We can think of this translation magnetic field as a non-vanishing strain configuration between the $z_1$ and $z_2$ planes that immediately relaxes back to the unstrained lattice outside this interval. This configuration induces opposite dislocation flux densities at the boundaries of the strained region. We schematically depict this geometry in Fig.~\ref{fig:grain_boundary}, where the red (blue) plaquettes contain positive (negative) dislocation fluxes, and the black arrows indicate hopping terms to which we apply the momentum-dependent Peierls phases.

Because we are interested in a surface response, we must use a layer-resolved Berry curvature to calculate the response coefficient $\mathcal{P}_\alpha$ for just the top (or bottom) surface of the system. The layer-resolved Berry curvature can be obtained by combining the projector onto the occupied subspace, defined as
\begin{equation}
    P(\vb{k}) = \sum_{\epsilon_i(\vb{k})<0}\ket{u_i(\vb{k})}\bra{u_i(\vb{k})}
\end{equation}
where $H(\vb{k}) \ket{u_i(\vb{k})} = \epsilon_i(\vb{k})\ket{u_i(\vb{k})}$, and the projector onto the $n^{\mathrm{th}}$ layer of the lattice, $P_n$, via the formula~\cite{essin_magnetoelectric_2009}
\begin{equation}
    \mathcal{F}^{ab}_n(\vb{k}) = \mathrm{Tr}\left[P(\vb{k}) \partial_{k_i} P(\vb{k}) P_n \partial_{k_j} P(\vb{k})\right].
\end{equation}
Using this formalism, the surface response coefficient is given by the momentum-space dipole moment of the layer-resolved Berry curvature summed over half the sites in the open direction:
\begin{equation}
    \mathcal{P}_x=\frac{1}{\pi}\sum_{n=1}^{N_y/2}\int_{\mathrm{BZ}}dk_xdk_z\, k_x\mathcal{F}^{xz}_n(k_x,k_z).\label{eq:surfBerryCurvature}
\end{equation}
The momentum and charge bound to the surface by dislocations and magnetic flux are calculated in a similar manner, i.e., layer contributions are summed over half of the sites in the open direction,
\begin{equation}
    j^0 = \sum_{n=1}^{N_y/2} j^0(n) ,\quad \mathcal{J}_x^0 = \sum_{n=1}^{N_y/2} \mathcal{J}_x^0(n).
\end{equation}

After carrying out these calculations, we show the $x$-momentum density bound to a strip of magnetic flux lines as a function of the magnetic flux in Fig.~\ref{fig:magnetic_surface} and plot the charge density bound to a strip of dislocations as a function of translation flux in Fig.~\ref{fig:dislocation_surface}. The momentum density bound to magnetic flux is linear in the magnetic flux with the correct proportionality constant $e\mathcal{P}_x/4\pi$. Here the value of $\mathcal{P}_x$ is determined by directly calculating Eq. \ref{eq:surfBerryCurvature}, which determines the slopes of the dashed lines in Figs. \ref{fig:magnetic_surface} and \ref{fig:dislocation_surface}. For a small dislocation flux value, $\Phi^T=1$, the charge density bound to dislocations matches the prediction of the effective action, but this relation becomes non-linear at higher values of $\Phi^T$ because of stronger lattice effects. As mentioned above for the bulk calculations, the open circles in Fig. \ref{fig:dislocation_surface} represent non-integral dislocation fluxes that are mathematically obtainable via our momentum-dependent Peierls factors, though their physical interpretation as a lattice defect is not clear.

\section{Momentum-Weighted Quadrupole Moment}
\label{sec:momentum_quadrupole_moment}
As one final physical phenomenon associated to our system of a quadrupole of higher order Weyl nodes, let us consider what is happening at the hinges. Because some regions of momentum-space harbor higher order topology in our model, we expect to find hinge modes and/or fractional charge per unit length along the hinge. Indeed, the hinge phenomena in our system are associated with the momentum planes that harbor a 2D QI. At half-filling, the sign of the electric quadrupole moment of these planes is ambiguous when the symmetries protecting the topology are enforced, i.e., the value $q_{xy} = e/2$ is equivalent to $q_{xy} = -e/2$. In the case of the QI phases of $H(\vb{k};k_x)$ and $H(\vb{k};k_y)$, the relevant quantizing symmetries are the pair of mirror times time-reversal symmetries. To choose the sign of the quadrupole moment we want to weakly break both of these symmetries, but preserve the product, i.e., preserve $C_2$ symmetry so that no electric dipole moment is allowed. Operationally, for a system with open boundaries, the symmetry breaking provides a prescription of how to fill the low-energy hinge states that is consistent with the sign of the quadrupole moment. Interestingly, our model has two distinct possible choices of symmetry breaking that we discuss below.

One possible choice of symmetry breaking is the perturbations $H'(\vb{k}) = \delta\sin(k_{x/y})\Gamma_0,$ which accomplish the required symmetry breaking for $H(\vb{k};k_{x/y})$ respectively. Since $\Gamma_0$ is odd under time-reversal, this term is time-reversal invariant, but it breaks both mirror symmetries since $\sin(k_{x/y})$ is odd under mirror $M_{x/y}$. As a consequence, this perturbation endows the positive- and negative-momentum intervals of the QI phases with the same sign of quadrupole moment. Hence the positive and negative momentum intervals add together to yield a finite bulk electric quadrupole moment (per $xz$ cross sectional area),
 \begin{equation}
    \begin{aligned}
        Q^{\mathrm{bulk}}_{xz} &= L_y\int\frac{dk_y}{2\pi} q_{xz}(k_y) \\
        &=\pm\frac{e L_y}{2\pi}\left(k_{\text{Dirac}} - k_{\text{Weyl}}\right),
    \end{aligned}
\end{equation}
where the sign is determined by the sign of $\delta$. The analogous quantity in the $yz$ plane is defined as
\begin{equation}
    \begin{aligned}
        Q_{yz}^{\text{bulk}} &= L_x \int\frac{dk_x}{2\pi} q_{yz}(k_x) \\
        &= \mp \frac{e L_x}{2\pi} \left(k_{\text{Dirac}} - k_{\text{Weyl}}\right).
    \end{aligned}
\end{equation}
The magnitude of the bulk electric quadrupole is determined solely by the separation between the projections onto the hinge BZ of the bulk Weyl nodes and surface Dirac nodes, $k_{\mathrm{Weyl}}$ and $k_{\mathrm{Dirac}}$, as these control the portion of the BZ that is occupied by the QI phase. 
 
\begin{figure}
\centering
\includegraphics[width=0.45\textwidth]{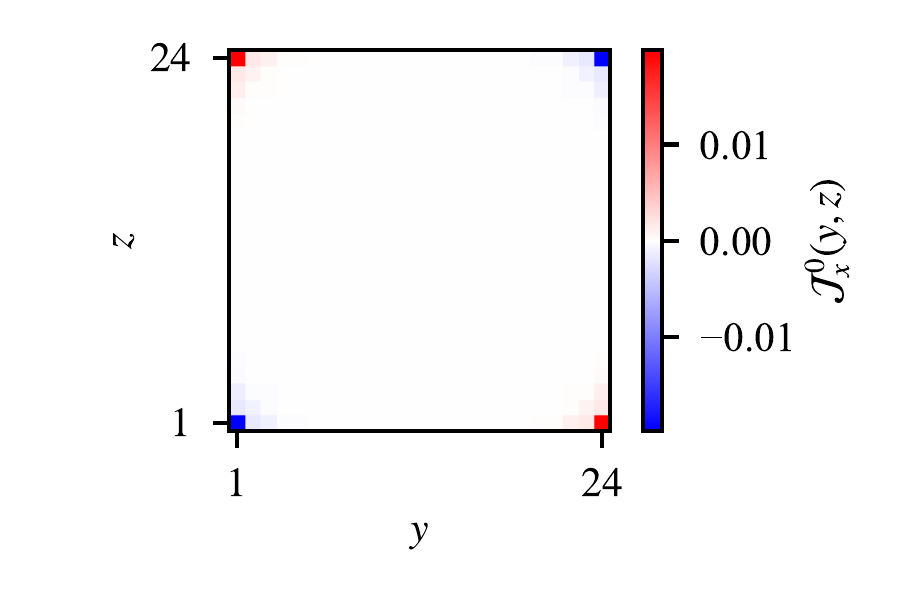}
	\caption{The position-resolved momentum density of $H(\vb{k}) + H''(\vb{k})$ with open boundary conditions along the $y$- and $z$- directions, $m=-0.3$, $\beta=-0.7$, $\gamma=0.5$, and $\delta=10^{-4}$.}
	\label{fig:momentum_quadrupole_density}
\end{figure}

Next we consider a second possible symmetry breaking perturbation $H''(\vb{k}) = \delta\Gamma_0.$ This term preserves the mirror symmetries but breaks time-reversal symmetry. It has the effect of endowing the positive- and negative-momentum intervals of QI phases with \emph{opposite} quadrupole moments. The resulting bulk electric quadrupole moment vanishes, since it receives equal and opposite contributions from each momentum interval. Instead the system realizes quadrupole moments of \emph{crystal momentum} density that have not previously been considered,
\begin{equation}
    \begin{gathered}
        K^y_{xz} = \frac{L_y}{e}\int\frac{dk_y}{2\pi} k_y q_{xz}(k_y) , \\
        K^x_{yz} = \frac{L_x}{e}\int\frac{dk_x}{2\pi} k_x q_{yz}(k_x).
    \end{gathered}
\end{equation}
The bulk crystal momentum quadrupole moment density manifests as momentum density bound to hinges, as shown in Fig.~\ref{fig:momentum_quadrupole_density}, where the momentum points along the hinges. Similar to the bulk electric quadrupole moment, the magnitude of the bulk crystal-momentum quadrupole moment is determined by the locations of the bulk Weyl and surface Dirac nodes,
\begin{equation}
    K^y_{xz} = \pm\frac{L_y}{4\pi}\left(k_{\text{Dirac}}^2 - k_{\text{Weyl}}^2\right),
\end{equation}
where the overall sign is again determined by the sign of $\delta$. It is interesting to note that this quantity can be concisely expressed in terms of the Weyl quadrupole moment $\bar{Q}$ and the surface Dirac dipole moment $\mathcal{P}_{x}$,
\begin{equation}
    \label{eq:k_xyz_p_q}
    K^y_{xz} = \pm \frac{L_y}{\pi}\left(\mathcal{P}_y^2 - 2 \bar{Q}\right),
\end{equation}
and therefore acts as a link between the bulk and surface mixed crystalline-electromagnetic responses. This is analogous to the response of higher order Weyl dipole systems, in which the extent of the Fermi arcs on the surface and the arcs on the hinge must satisfy a sum rule~\cite{ghorashiHigherOrderWeylSemimetals2020}.

We note that the bulk quadrupole moment of crystal momentum density is well-defined only when the bulk electric quadrupole moment vanishes, as its value can otherwise be arbitrarily changed by shifts of the BZ origin $\mathbf{k} \rightarrow \mathbf{k} + \mathbf{k}'$. This is exactly what happens when we choose the $H''$ perturbation since the total bulk quadrupole moment vanishes. The invariance of the bulk quadrupole moment of crystal momentum under shifts of the BZ can also be seen from the definition in Eq.~\eqref{eq:k_xyz_p_q}. The Weyl quadrupole moments $Q_{xx}$ and $Q_{yy}$ are invariant under such shifts because the $C_{2z}$ symmetry of the Hamiltonian forces the Weyl dipole moments in the $k_x$-$k_y$ plane to vanish, and the surface Dirac dipole moments $\mathcal{P}_{x/y}$ are invariant because the product of the $M_{1,\pm1}$ and $C_4M_z$ symmetries form a surface $M_z$ symmetry that forces the surface Chern number to vanish.

\section{Conclusion}
\label{sec:conclusion}
In this work we made the first steps towards understanding the interplay between higher order topology and mixed crystalline-electromagnetic responses. By constructing and analyzing an explicit model, we showed that elevating a quadrupole arrangement of Weyl nodes, which is known to exhibit a bulk mixed crystalline-electromagnetic response, to higher order Weyl nodes produces an additional mixed crystalline-electromagnetic \emph{surface} response. We further demonstrated that the surface response originates from the higher order QI phases of the Hamiltonian in the foliated BZ. We additionally found that adding symmetry breaking perturbations can produce bulk quadrupole moments of either electric charge or crystal momentum, depending on the particular perturbation chosen.

These results motivate a number of different directions for future research. Of primary importance is identifying promising material platforms in which these mixed crystalline-electromagnetic responses can be observed. The response we predict in this work requires the system to possess both a bulk Weyl quadrupole moment and a surface Dirac dipole moment. As for the crystal symmetry ingredients, for the Weyl quadrupole moment to be well defined, the Weyl dipole moments in the plane of the quadrupole must vanish, which can be guaranteed by mirror symmetry or a set of $C_2$ symmetries (time reversal symmetry would also suffice,  although that would prevent observation of the momentum quadrupole). The surface Dirac dipole moment similarly requires the surface Chern number to be zero, which can be enforced by the presence of a surface mirror symmetry or a time reversal symmetry (as long as the bulk is not a 3D topological insulator). These symmetries, along with the possible breaking of TRS either by magnetic ordering or an applied magnetic field, are necessary to observe the mixed crystalline-electromagnetic response. Combining these symmetry requirements with the tools provided by topological quantum chemistry may provide a route to identifying materials that host this mixed crystalline-electromagnetic response~\cite{bradlyn2017topological, elcoro2021magnetic}. 

There are a number of systems that are likely to host similar types of mixed responses and warrant further study. Higher order analogs of two-dimensional Dirac quadrupole semimetals and three-dimensional nodal line semimetals~\cite{dubinkin_in_prep} are particularly promising, as are higher order nodal superconductors~\cite{ghorashi2019second, PhysRevB.103.184510, wuNodalHigherorderTopological2022a, simonHigherorderTopologicalSemimetals2022} and higher order non-Hermitian TSMs~\cite{ghorashi2021nonDirac,ghorashi2021nonWeyl,liu2021higher, bidNonHermitianHigherOrderWeyl2022}. Furthermore, there are promising metamaterial platforms in which one could generate our model. Both Weyl points~\cite{lu2015experimental} and higher order quadrupole topology~\cite{serra2018observation,peterson2018quantized,imhof2018topolectrical} have each been demonstrated separately in experiment, so combining the two is plausibly achievable. In these systems it may even be possible to extract information about the crystal momentum, as was recently accomplished in a topoelectric circuit experiment studying higher rank surface states~\cite{zhu2023higher}.

Interestingly, our model also presents a platform in which to study quantum oscillations, as the combination of surface Fermi arcs and zero-energy hinge arcs may provide unusual circuits for electrons to traverse~\cite{potter2014quantum}. The properties of these systems in strong magnetic fields may also be a fruitful line of pursuit as the zeroth Landau level of the bulk Weyl nodes must coordinate with the zeroth Landau level of the surface Dirac fermions. We leave these studies to future work. 

\section*{Acknowledgements}
    We thank Oleg Dubinkin and Julian May-Mann for helpful discussions. We acknowledge the NSF-Supported UIUC REU program under award numbers PHY-1950744 and PHY-2244433. TLH and MH thank ARO MURI W911NF2020166 for support.


%

\end{document}